\documentclass[12pt]{article}
\usepackage{amsmath}
\usepackage{amssymb}
\usepackage{latexsym}
\usepackage{graphicx}

\newcommand{\fab}{F_{\alpha \beta}}
\newcommand{\fmn}{F^{\mu \nu}}
\newcommand{\varep}{\varepsilon^{\nu \alpha \beta}}
\newcommand{\astf}{{^\ast F}}
\newcommand{\calD}{\mathcal{D}}

\numberwithin{equation}{section}

\begin{document}
\title{Fifty Years of Yang-Mills Theory\\ and my Contribution to it.}
\author{R. Jackiw \footnote{Email: jackiw@lns.mit.edu}\\
{\small\itshape Center for Theoretical Physics}\\
{\small\itshape Massachusetts Institute of Technology}\\
{\small\itshape Cambridge, MA 02139-4307}}
\date{MIT-CTP-3484}
\maketitle

\thispagestyle{empty}

\begin{abstract}
\noindent On the fiftieth anniversary of Yang-Mills theory, I review the contribution
to its understanding by my collaborators and me.\\ 
Contents: 1.Gauge Theories and Quantum Anomalies; 
2.Mathematical Connections; 
3. Gauge Field Dynamics other than Yang-Mills;
4. Gauge Formalism for General Relativity Variables;
A. Christoffel connection as a gauge potential,
B. Gravitational Chern-Simons term from gauge theory Chern-Simons term,
C. Coordinate transformations in general relativity and gauge theory,
(i) Response to changes in coordinates\\
(ii) Invariant fields and constants of motion. References.
\end{abstract}
To be published by World Scientific.
\newpage
\thispagestyle{empty}
\mbox{}
\newpage
Following some precursors (Klein, Pauli, Shaw), Yang and Mills invented non-Abelian gauge theory
half a century ago [1]. Governed by the Yang-Mills Lagrange density 
\[
\mathcal{L}_{YM} = \frac{1}{2} <F^{\mu \nu} F_{\mu \nu}>,
\]
where $F_{\mu \nu}$ is the Lie algebra/matrix valued gauge field
strength constructed from a gauge potential $A_\mu$
\[
F_{\mu \nu} = \partial_\mu A_\nu - \partial_\nu A_\mu + [A_\mu, A_\nu],
\]
the model generalizes in a natural and elegant fashion Maxwell electrodynamics, to
which it also reduces in the Abelian case. (Brackets $<>$ denote matrix trace.)

Twenty years passed before physicists learned how to quantize, renormalize
and put the theory to phenomenological use describing the dynamics of fundamental
elementary particles -- a length of time comparable to the interval between the
invention of quantum physics by Planck and its final formulation by Heisenberg and
Schr\"{o}dinger. In the early seventies, the work of our editor `tHooft [2] and his
teacher Veltman [2] made it possible to perform calculations for strong, weak and electromagnetic
interactions, based on well-defined, non-Abelian gauge theory
models, which became subsumed in the ``standard model" for elementary
particles. While important investigations unraveled the novel dynamics
(confinement, renormalization group and asymptotic freedom, large-N limit), my
collaborative research focused on the kinematical properties of non-Abelian gauge
fields; properties that become visible when close attention is paid to the gauge
theory's mathematical (geometrical, topological) structures, which nevertheless
affect physical content. In this celebratory review, I shall first summarize some of our
work, and then present remarks on gauge theoretic aspects of gravity theory.

\section{Gauge Theory and Quantum Anomalies}
A precursor to much mathematical analysis of gauge theories is the chiral anomaly, discussed by Bell and me [3] (also Adler and, earlier, others [4]) before non-Abelian theories entered center-stage. We showed that a chiral symmetry of classical dynamics does not in general survive quantization: In the presence of fermions, the continuity equation for the classically conserved, but quantum
mechanically non-conserved chiral current, $J^\mu_A$, becomes proportional to the ``anomaly"
after quantization. 
\begin{equation}
\partial_\mu J^\mu_A \propto \ <{^\ast \negthickspace F}^{\mu \nu} F_{\mu \nu}>
\label{one1}
\end{equation}
($^\ast F^{\mu \nu}$ is the dual of $F_{\mu \nu}: {^\ast F}^{\mu \nu} \equiv \frac{1}{2} \varepsilon^{\mu \nu \alpha \beta} F_{\alpha \beta}$.) 
Physicists later recognized the quantity on the right as the Chern-Pontryagin
topological density. The anomaly allowed evading a chiral symmetry-based,
model-independent no-go theorem prohibiting the neutral pion from decaying into
two photons -- a process seen in Nature. Evidently quantum effects destroy the
apparent (classical) chiral symmetry and negate the physically unacceptable
prohibition. This was very much appreciated by Bell's good friend Veltman, who
(with Sutherland) established the no-go result [5]. His pupil `tHooft later made
important connections between the chiral anomaly and properties of the standard
model (see below).

The first application of the chiral anomaly to the standard model came with the observation by Gross and me [6] (also Bouchiat,
Iliopoulos and Meyer [7]) that the `tHooft-Veltman argument for renormalizability
of gauge theories remains valid only if fermion content is arranged so that gauge
fields couple to currents that are free of anomalies, in which case the Yang-Mills
equation with sources $J^\mu$ is self-consistent.
\begin{eqnarray}
&& \partial_\mu F^{\mu \nu} + [A_\mu, F^{\mu \nu}] \equiv \mathcal{D}_\mu F^{\mu \nu} = J^\nu \nonumber\\
0 &=& \mathcal{D}_\nu \mathcal{D}_\mu F^{\mu \nu} = \mathcal{D}_\nu J^\nu
\label{one2}
\end{eqnarray}

This requirement together with the strength of the anomaly, fixed experimentally by the $\pi^0 \to 2 \gamma$ decay amplitude, provides up to now one of the few principles for determining the color and family structure of elementary fermions (quarks, leptons) in Nature.

Nevertheless, the subject lay fallow until instantons were found by Belavin, Polyakov,
Schwartz and Tyupkin \cite{abelvin} (who called them ``pseudoparticles").  
Instantons are classical gauge field configurations in imaginary time (Euclidean
space-time) that are self-dual or anti self-dual.
\begin{equation}
^\ast F^{\mu \nu} = \pm F^{\mu \nu}
\label{one3}
\end{equation}
Therefore, they satisfy the classical, imaginary-time, sourceless Yang-Mills equation,
\begin{equation}
 \mathcal{D}_\mu F^{\mu \nu} =0,
\label{one4}
\end{equation}
by virtue of the Bianchi identity for the dual field strength.
\begin{equation}
\mathcal{D}_\mu {^\ast F^{\mu \nu}} = 0
\label{one5}
\end{equation}
Moreover, the 4-dimensional integral of the Chern-Pontryagin density evaluated on instanton configurations takes values fixed by an integer that labels the homotopy class to which the gauge field belongs. The non-vanishing value for this integral came as a surprise, because the Chern-Pontryagin density is itself a total divergence, so its integral is a surface term frequently ignored by physicists.
\begin{eqnarray}
\frac{1}{4}<{^\ast F^{\mu \nu} F_{\mu \nu}}> = \partial_\mu K^\mu \qquad \qquad
\qquad
\nonumber\\ K^\mu \equiv \varepsilon^{\mu \alpha \beta \gamma} <\frac{1}{2}
A_\alpha \partial_\beta A_\gamma + \frac{1}{3} A_\alpha A_\beta A_\gamma>
\label{one6}
\end{eqnarray}

$K^\mu$ is called the Chern-Simons or anomaly current, whose time component
$K^0$ contains only spatial quantities. $K^0$ is also called the Chern-Simons density,
about which I shall have more to say later.
\begin{equation}
CS(A) \equiv K^0 = \varepsilon^{ijk} < \frac{1}{2} A_i \partial_j A_k + \frac{1}{3} A_i A_j A_k>
\label{one7}
\end{equation}

The surprising relevance of all this to quantum physics in Minkowski space-time comes about for the following reasons. Since the baryon number current in the standard model does not couple to a gauge field, it can remain anomalous, even though it is a vector quantity with no chiral component. This possibility was appreciated already in the original paper with Bell [3], but no physical consequence was drawn, because the total
accumulated
 change of the anomalous charge, $Q_A = \int d^3 x J^0_A$, is a surface term of no
apparent importance before the advent of instantons.
\begin{equation}
\triangle Q_A \equiv \int^{\infty}_{-\infty} d t \frac{d}{dt} \int d^3 x J^0_A \propto \frac{1}{4} \int dt d^3 x <{^\ast F^{\mu \nu}} F_{\mu \nu}> 
= \int d t d^3 x \partial_\mu K^\mu \qquad \qquad
\label{one8}
\end{equation}

However, `tHooft realized that this reasoning is inadequate [9]. He observed that instantons can be used to evaluate approximately the Yang-Mills functional integral continued to imaginary time. Since also the instanton-dominated integral of the Chern-Pontryagin density is non-vanishing, `tHooft concluded that baryon number is not conserved in the standard model. By evaluating the Euclidean functional integral in a Gaussian approximation around the instanton solution of Belavin {\it et al.}, he calculated the baryon lifetime. Fortunately it is exponentially small, but diamonds
in principle are not forever.

There remained much to be done in extending `tHooft's results. One wished to identify the physical mechanism in Minkowski space-time behind the approximate evaluation of an Euclidean functional integral. Also `tHooft found that his answer depends on an angle $\theta$, which is not seen in gauge field dynamics; again a physical explanation was needed.

By recalling known procedures in condensed matter physics and quantum chemistry Rebbi and I [10]
(also Callan, Dashen and Gross [11]) explained that classical paths in imaginary time signal quantum
tunneling, whose probability amplitude in the semi-classical approximations is given by the Euclidean
functional integral in Gaussian approximation around the imaginary time path. But where does the
tunneling take place? To answer this, and also to understand the
$\theta$-angle, we examined the gauge theory in the Schr\"{o}dinger representation, where quantum states
are described by wave  functionals $\Psi ({\bf A})$ defined on configuration
space variables -- here the spatial vector potentials $\bf A$. (Although the reality of
a space of gauge field variables is obscure, it is no more obscure than the 3
N-dimensional space on which N-body, $N>1$, quantum mechanical wave functions
are defined.)

We then drew the following qualitative but exact picture of the quantum field theory
\cite{crebbi}. The Gauss law and conventional gauge fixing (of the kind described by
Faddeev in this volume) ensure that $\Psi ({\bf A})$ is unchanged when $\bf A$
undergoes a gauge transformation that is deformable to the identity; so called a
``small'' gauge transformation. However, in non-Abelian groups there are gauge
transformations that are homotopically non-trivial and cannot be connected to the
identity. For these ``large'' transformations, labeled by an integer
$n$ that indexes the homotopy class to which the gauge transformation belongs, $\Psi
({\bf A})$ changes by a phase. This phase is the $\theta$-angle encountered by
`tHooft in his calculation. 

In the classical ground state, the magnetic field $B^i \equiv {^\ast \negthinspace F}^{i0}$ vanishes. This is achieved not only with vanishing $\bf A$, but also with pure-gauge vector potentials. Since large gauge functions fall into classes labeled by the integers, the potential energy profile takes on a periodic structure, reminiscent of a Bloch crystal
and depicted in the Figure below. \vspace{12pt}

\begin{figure}[h]
\begin{center}
\includegraphics{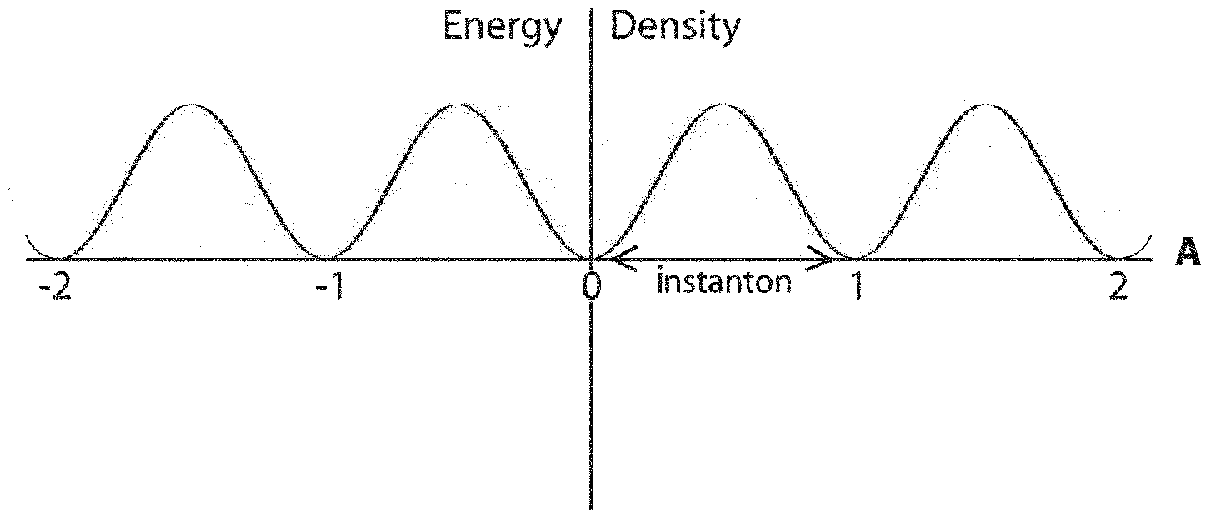}
\end{center}
\end{figure}
\vspace{-8pt}
The zero-energy troughs correspond to pure gauge vector potentials constructed from gauge functions belonging to the $n^{th}$ homotopy class. Thus the classical ground state is (infinitely) degenerate. Quantum mechanical tunneling lifts this degeneracy and creates energy bands. Usually in quantum field theory tunneling is suppressed by infinite energy barriers between degenerate vacua -- this leads to spontaneous symmetry breaking. However, in Yang-Mills theory there are
imaginary time paths in field space, which avoid infinite barriers and interpolate
between the vacua. Semi-classically these are the instantons., as can be verified by
examining their large (imaginary) time asymptotes. The final picture is portrayed
schematically in the above Figure. 

To go beyond qualitative considerations we constructed an explicit functional
of $\bf A$, which is invariant against small gauge transformations but not large ones.
\cite{crebbi} This is just the spatial integral of the Chern-Simons density, introduced
in (\ref{one7}). One readily checks that the Chern-Simons term
\begin{equation}
W({\bf A}) = \frac{1}{4\pi^2} \int d^3 x CS({\bf A}) = \frac{1}{4 \pi^2} \int d^3 x
\varepsilon^{ijk} <\frac{1}{2} A_i \ \partial_j A_k + \frac{1}{3} A_i \ A_j \ A_k>,
\label{one9}
\end{equation}
satisfies the Gauss law by virtue of 
\begin{subequations}
\begin{eqnarray}
{\delta W ({\bf A})} = \frac{1}{4\pi^2}\ \negthinspace \int d^3 x
\delta {\bf A} \cdot {\bf B}, \label{one10a} \\  
{\bf \mathcal{D}} \cdot {\bf B} = 0.
\label{one10b}
\end{eqnarray}
\end{subequations}
But when $\bf A$ is gauge transformed by a large gauge function in the $n^{th}$ homotopy class,
$W({\bf A})$ shifts by the ``winding'' number $n$.
\begin{eqnarray}
W ({\bf A}^g) &=& W ({\bf A}) - \frac{1}{24 \pi^2} \int d^3 x \varepsilon^{ijk} < g^{-1}
\partial_i g\ g^{-1} \ \partial_j g\ g^{-1} \partial_k g > \label{one11}\\
A^g_i &\equiv& g^{-1} A_i \ g + g^{-1} \partial_i \ g
\label{one12}
\end{eqnarray}
The last term in (\ref{one11}) evaluates the winding number of $g$ (for well-behaved $g$).

Every Yang-Mills wave functional can be presented as
\begin{equation}
\Psi ({\bf A}) = e^{i\theta W({\bf A})} \Psi_{inv} ({\bf A}),
\label{one13}
\end{equation}
where $\Psi_{inv} ({\bf A})$, is invariant against all gauge transformations, small and large, while the non-invariance of $\Psi({\bf A})$ is contained in the universal phase involving $W({\bf A})$. 
\begin{equation}
\Psi ({\bf A}) \to e^{i n \theta} \Psi ({\bf A})
\label{one14}
\end{equation}
But in quantum theory a universal phase of wave functions may be removed at the expense of adding the
time derivative of the phase to the theory's Lagrangian. When this procedure is
carried out with the help of (\ref{one6}) and (\ref{one7}) for the problem at hand, the Yang-Mills
quantum theory becomes equivalently described by completely gauge invariant
states, $\Psi_{inv} ({\bf A})$, while the Yang-Mills Lagrangian acquires the addition
$\theta \int d^3 x {^\ast\negthinspace F^{\mu \nu}} F_{\mu \nu}$, which does not
contribute to equations of motion because it is a total time derivative. This shows
that the $\theta$-angle is associated with Lorentz invariant but CP (or T) violating
phenomena, which through our analysis are understood in exact terms, while
instantons provide a semi-classical description. We are left with a puzzle: What fixes
the magnitude of $\theta$, whose experimental consequences ({\it e.g.}  neutron
electric dipole moment) have never been seen? (An analogy with the cosmological
constant puzzle is apparent.)

\section{Mathematical Connections}
The anomaly-based instanton investigation of the standard model did not produce any useful numbers for experimentalists to measure. But it affected deeply our understanding of the theory. Also it suggested a wealth of interesting mathematical problems to which Rebbi and I found solutions by methods drawn from analysis, geometry and topology.

We proved that the Belavin {\it et al.} instanton preserves an $SO(5)$ symmetry
subgroup of the $SO(5,1)$ conformal invariance group for Euclidean Yang-Mills
theory
\cite{rjebbi}. This allows a group theoretical classification of motions in the presence of
the instanton. Furthermore, use of $SO(5)$ covariant coordinates yields simple and
elegant formulas, so that evaluating the functional integral in a Gaussian
approximation around the instanton becomes transparent. We \cite{rjebbi2} (also
Schwartz, as well as Atiyah, Hitchin and Singer \cite{aschwartz})  showed that the
most general instanton configuration with Chern-Pontryagin index 
$ (\equiv \int d^4 x <{^\ast F}^{\mu \nu} F_{\mu \nu}>) \propto  n$ can be viewed as a
non-linear superposition of $|n|$ individual instantons, which depends on $8
|n|$ parameters for the $SU(2)$ group: $4|n|$ positions in 4-dimensional space, $|n|$
instanton sizes, and $(k^2-1) |n|$ group variables of $SU(2)$. Also following a
suggestion by `tHooft, we exhibited the most general, explicit multi-instanton 
formula. \cite{cnohl} Our expression is closed under conformal transformations and
maximizes the parameter count for
$|n| =1\ \mbox{and}\ 2$. (For $|n|>2$, no explicit formula for the general solution is
known, but a procedure for constructing it at given $|n|$ has been found by Atiyah,
Drinfeld, Hitchin and Manin. \cite{aityah})

Much of this analysis involves zero eigenvalue solutions to 4-dimensional elliptic differential equations in the
presence of instantons. The number of these zero modes is determined by the Chern-Pontryagin index of the
background gauge field. This is the statement of the celebrated Atiyah-Singer index theorem which appeared
in physics for the first time because of these considerations. \cite{spinor} Indeed the
anomaly equation (1.1) may be viewed as a local version of the theorem.

The zero modes arise not only with 4-dimensional instantons, but also in the presence of extended, topologically interesting field configurations in other
dimensions. For example, `tHooft and Polyakov
\cite{magmono} found that magnetic monopole configurations are present as classical
static solutions to some gauge theories based on semi-simple groups, like
$SU(2)$. Rebbi and I \cite{isospin}  (also Hansenfratz and `tHooft \cite{hasenfratz})
then  showed that spinless boson fields, carrying half-integer isospin, bind to the
monopole with zero energy and form a spin
$\frac{1}{2}$ excitation in the quantum theory, even though all ``constituents" carry
integer spin. While it is uncertain whether magnetic monopoles and the associated
spin fractionization have a physical presence in Nature (they are not features of the
standard model) analogous effects can arise in condensed matter systems that are
available in a laboratory. For example, 1-dimensional topological kinks form domain
walls in lineal polymers like polyactelene. Electrons propagating across these kinks
experience fractionization of their quantum numbers -- an effect that has been
observed experimentally. \cite{soliton21}

The zero modes associated with monopoles and kinks arise in elliptic differential equations in odd dimensions, where the Atiyah-Singer index cannot be used because it is restricted to even dimensions.
Callias, a student at that time, provided the necessary extension, and the ``Callias index" is now used for counting zero modes in odd-dimensional
spaces. \cite{ccallias}

The mathematical activity surrounding instantons and other extended objects, like monopoles, vortices and
kinks, seeded an interaction between physics and mathematics, which is still
flourishing. At an American Physical Society meeting, where I summarized the
above results
\cite{rjackiw}, Singer declaimed a poetic pean to physics -- mathematics
collaboration:\\
\vspace{-5pt}

\begin{minipage}{6.5in}
\quad \quad  \begin{minipage}{2in} 
{\it In this day and age\\
The physicist sage\\
Writes page after page\\
On the current rage\\
The gauge}
\end{minipage}
\
\begin{minipage}{2in}
{\it Mathematicians so blind\\
Follow slowly behind\\
With their clever minds\\
A theorem they'll find\\
Only written and signed}
\end{minipage}
\
\begin{minipage}{2in}
{\it But gauges have flaws\\
God hems and haws\\
As the curtain He draws\\
O'er His physical laws\\
It may be a lost cause}
\end{minipage}
\end{minipage}
{\hbox{\rightline{I. Singer}}} 
\\ [8pt]
\vspace{-2pt}
The interaction with mathematics became fueled anew when physicists called attention to further gauge theoretic
invariants.
\section{Gauge Field Dynamics other than Yang-Mills}
Exploration of Yang-Mills theory brought physicists' attention to gauge theoretic
invariants, other than the Yang-Mills Lagrange density $\frac{1}{2} <F^{\mu \nu}
F_{\mu \nu}>$. I have already discussed the anomaly-determined Chern-Pontryagin
density
$<{^\ast F}^{\mu\nu} F_{\mu \nu}>$ (1.1). It does not contribute to equations of motion 
because it is a total derivative, but it controls the $\theta$-angle. Further there is the
Chern-Simons current $K^\mu$ (1.6) and density $K^0 \equiv CS(A)$ (1.7) --
``anti-derivatives" of the Chern-Pontryagin density. Defined on 3-space, $CS(A)$ is not
gauge invariant. But its properly normalized integral  over 3-space -- the
Chern-Simons term $W({\bf A})$ (1.9) -- responds only to large gauge transformatives
by the integer winding number of the gauge function. 

$W({\bf A})$ first entered physics as the phase of Yang-Mills quantum states, (1.13),
where it is responsible for the $\theta$-angle.

Another remarkable but futile role for $W({\bf A})$ is that $e ^{\pm\, 4 \pi^2 W({\bf
A})}$ solves the Yang-Mills functional Schr\"{o}dinger equation with zero eigenvalue
by virtue of (1.10a):
$H_{YM} (\frac{1}{i} \frac{\delta}{\delta {\bf A}}, {\bf A} )\ e^{\pm 4 \pi^2 \,W({\bf A})}
= 0$, where
$H_{YM}$ is the Yang-Mills Hamiltonian. It is astonishing that such a non-linear,
integro-differential, functional equation possesses a simple and explicit solution.
Unfortunately, $e^{\pm 4 \pi^2 W({\bf A})}$ cannot describe a physical state because
$| e^{\pm 4 \pi^2 W({\bf A})} |$ is not gauge invariant against large gauge
transformations, and grows exponentially for large (functional) argument. [This
``solution" is a great gauge theory teaser, but in fact it possesses a quantum
mechanical analog: The zero-eigenvalue Schr\"{o}dinger equation for a
2-dimensional, $(x, y)$, isotropic harmonic oscillator (unit frequency) is solved by
$e^{\pm x y}$, which diverges for large  $x$ or $y$, and has no place in the quantal
Hilbert space for this system.]

The integer valued gauge non-invariance of $W(A)$ does not prevent using that
3-dimensional quantity in the action for a gauge theory on a (2+1)-dimensional
space-time. Because its variation (1.10a) is gauge covariant, $W(A)$ contributes a
gauge covariant quantity to the equation of motion. Also in a quantum theory only
the phase exponential of the action need be gauge invariant. With this in mind, Deser
and I
\cite{3dmass} (together with students and postdoctoral fellows) proposed that $m
W(A)$ be included in the action for a novel (2+1)-dimensional gauge theory, where
$m$ is strength of the new term. [Here $A_\mu$ is a ($2+1$)-dimensional space-time
covariant vector potential, not a 3-dimensional spatial vector $\bf A$.]

In order to preserve gauge invariance of the phase exponential of the action for a
non-Abelian theory, $m$ must be quantitized to be an integral multiple of $2 \pi$.
Then
$e^{i m W(A)}$ remains invariant even against large gauge transformations, which shift
$W(A)$ by an integer.  This coupling constant quantization is the precise field theoretic analog of
Dirac's celebrated monopole strength quantization. Of course in an Abelian theory, there are no large gauge
transformations; $W(A)$ is gauge invariant; quantization of the interaction strength is
not needed.

The modified but gauge covariant equation of motion, with a covariantly conserved
source, reads
\begin{eqnarray}
\mathcal{D}_\mu \fmn + \frac{m}{8 \pi^2} \ \varep \fab = J^\nu, \label{three1}\\
-\varep \calD_\alpha \astf_\beta + \frac{m}{4 \pi^2} \, \astf^\nu = J^\nu.
\label{three2}
\end{eqnarray}
The non Yang-Mills term on the left comes from varying $m W(A)$, see (\ref{one10a}). Eq.
(\ref{three2}) exhibits the dual field $(\astf^\mu \equiv \frac{1}{2} \ \varepsilon^{\mu \alpha
\beta}
\fab)$, which in 3 dimensions is a vector, obeying the Bianchi identity.
\begin{equation}
\calD_\mu \astf^\mu =0
\label{three3}
\end{equation}
Note that (\ref{three2}) is consistent with (\ref{three3}).

The dimension of $m$ is mass (in units of the gauge coupling constant) and one sees
either from the linear portion of (\ref{three1}) or from the linear Abelian case
that gauge excitations in this theory are massive, while retaining gauge invariance!
This provides yet another example where gauge invariance does not enforce
masslessness for gauge field excitations. Note that $P$ and $T$ are violated by the
mass term.

Because the mass term in (\ref{three1}) has one fewer derivative than the usual Yang-Mills
kinetic term, it dominates at low energies and large distances. In the absence of the
Yang-Mills term, equation (\ref{three1}) reduces to a field-current identity.
\begin{equation}
\frac{m}{8\pi^2} \ \varep \fab = J^\nu
\label{three4}
\end{equation}
This is especially interesting in the Abelian case -- planar electrodynamics -- where the
components of (\ref{three4}) read
\begin{subequations}
\begin{eqnarray}
B = -\frac{4\pi^2}{m} \ \rho, \label{three5a}\\
E^i = \frac{8\pi^2}{m}\ \varepsilon^{ij} J^j.
\label{three5b}
\end{eqnarray}
\end{subequations}
Here $E^i$ is a planar electric field; $B$, a magnetic field perpendicular to the plane;
$\rho$ is a charge density and $J^i$ a planar current. A further consequence of
(\ref{three5a}), which also follows from the time component of (\ref{three1}) or (\ref{three2}), is the
integrated statement
\begin{equation}
N = -\frac{4\pi^2}{m} Q,
\label{three6}
\end{equation}
where $N$ is the magnetic flux through the plane and $Q$ is the charge.

Relations like (\ref{three5a})-(\ref{three6}) arise in descriptions of the quantum Hall regime, and the
Chern-Simons term has been widely used to model this planar effect. More recently
high temperature superconductivity gave impetus to applications of Chern-Simons
structures in speculative descriptions of that phenomenon. Also physics returned the
Chern-Simons term to mathematics when Witten used it in a functional integral
formula for knot invariants.

My investigations of Chern-Simons based gauge theories mostly concern the truncated
equations (\ref{three4}) with source currents constructed from specific field theoretic or point
particle variables. The dynamics of the sources is also self consistently included. 

For sources made from relativistic scalar fields with precisely tuned self interactions,
E. Weinberg and I
\cite{selfdual} (also Hong, Kim and Pac \cite{multivor}) found static vortex solutions,
similar to Ginsburg-Landau (Nielsen-Olesen) vortices at the boundary between type
I and type II superconductors.  Although equations are only partially integrable, we
showed that in the Abelian case both topological and non-topological vortices are
present. Furthermore, Pi and I \cite{pisoliton} considered non-relativistic scalar field
dynamics, more specifically the non-relativistic limit of the above mentioned model.
In that case the static equations are completely integrable, providing explicit profiles
for the vortices.

It is known that conventional vortex models, without the Chern-Simons term,
support only charge-neutral vortices. On the other hand, the Chern-Simons
vortices are charged, by virtue of (\ref{three6}). Also generically, they carry arbitrary,
unquantized angular momentum. This is a consequence of planar dynamics where
rotations are Abelian and angular momentum need not be quantized. \cite{reviews}

It is important that the Chern-Simons term is not merely an {\it
ad hoc}
$P$ and
$T$ violating addition to planar gauge theories, which could be included or omitted at
will. Even when it is absent in a bare Lagrangian containing fermions, it arises from
radiative corrections. Massless fermions in ($2+1$)-dimensional space-time
preserve planar parity invariance. Nevertheless they induce a parity violating
Chern-Simons term. This is the so called ``parity anomaly", discovered by
Redlich \cite{nonvariance}, a student during that research period. It is the
odd-dimensional analog of the even-dimensional chiral anomaly. (The fermion
determinant, which is responsible for this effect, can also be evaluated in a heat bath
environment at finite temperature; there its  response to large gauge
transformations is especially intricate.)

Both chiral and parity anomalies enforce quantum mechanical symmetry breaking in
theories that before quantization possess symmetries associated with masslessness.
One more instance of this phenomenon is known: anomalous quantum mechanical
breaking of scale and conformal invariance. Generically these do not survive in
non-trivial quantum field theories, not even in quantum mechanics. \cite{holstein}
Although not necessarily confined to gauge theories, nor possessing any significant
topological aspects, the scale/conformal anomalies share with the previous two
the property of destroying an enhancement of symmetry that masslessness would
entail. This common feature prepared Coleman and me to understand and explain
why the relevant currents are not conserved, with anomalous divergences
governed by the anomalously non-vanishing trace of the energy-momentum
tensor. \cite{dilatation} It appears that Nature abhors masslessness, but we do not
know why.

Finally we note that the Chern-Simons term $mW({\bf A})$, viewed as a quantity
defined on three spatial dimensions, can be inserted into the action for a Lorentz
invariant theory in ($3+1$) dimensional space-time. Due to the dimensional
mismatch, this acts as a source of Lorentz and CTP violation. The idea has been
developed for electrodynamics by Carroll, Field and me. \cite{parityvio} There it
produces a modification of the Maxwell equations only in Amp\`{e}res law, which in
the Lorentz violating theory reads
\begin{equation}
-\frac{\partial {\bf E}}{\partial {\bf t}} + {\boldsymbol \nabla} \times {\bf B} = {\bf J}
+\frac{m}{4\pi^2}\, {\bf B}.
\label{three7}
\end{equation}
(A source current containing a contribution from a magnetic field is familiar in
magnetohydrodynamics.) The physical consequence of (\ref{three7}) is that the vacuum
becomes birefringent, and propagating light waves undergo a Faraday-like rotation.
Light from distant galaxies provides an experimental measure of this effect. Available
observational data indicates that it does not occur in Nature; $m=0$.

\section{Gauge Formalism for General Relativity Variables}
 General relativity with its diffeomorphism invariance embodies the symmetry
of local translations. Therefore, one should try presenting the theory in a
formalism similar to that of a gauge theory. This has been achieved in lower
dimensions.
 
 For 3-dimensional Einstein theory in the {\it Dreibein}/spin connection formulation, field variables are
gauge potentials for the local Poincar\'{e} or deSitter groups: local translations are gauged by {\it
Dreibeine}, local Lorentz rotations, by spin connections. Dynamics is not of the
Yang-Mills form; rather the Lagrange density is a Chern-Simons term based on the
Poincar\'{e} or deSitter groups.
 Einstein theory does not exist in two dimensions, but various ``dilaton" gravities can
be formulated as gauge theories based on $SO$ (2.1) or (extended) Poincar\'{e}
groups, with dynamics governed by a ``BF" Lagrange density. In neither
dimensionality do these models support propagating, dynamical degrees of freedom.
 
 The above successes do not extend to four dimensions, where no complete gauge
theoretic formulation has been found for Einstein's theory. The closest we have again
uses {\it Vierbeine}/spin connections, with the latter gauging the local Lorentz group,
while the {\it Vierbeine} remain as covariant, additional variables.
 
 However, if we put aside the issue of gravitational dynamics and focus only on the gravitational field variables, we can find many
(notational) analogies to gauge fields. These analogies are useful for motivating and constructing
gravitational counterparts to gauge theoretic entities. Correspondingly, aspects of
general relativity can inform topics in gauge theory.
 
 Here I shall provide a dictionary between gauge theoretic and general relativistic variables, and then
use the relationship between them for further constructions both in general
relativity and gauge theory. It is likely that the gravity-gauge theory connection
described here is familiar to some (for example Bardeen and
Zumino \cite{gravtheor}) but I know of no text book discussion. Thus I hope that my
presentation will lead to wider appreciation of these useful formulas.

 \def\theequation{4.A.\arabic{equation}}
 \subsection*{A. Christoffel connection as a gauge potential}
 Consider the Christoffel connection $\Gamma^{\mu}_{\alpha \nu}$ (in any dimension $d$) and view it
as the ($\mu, \nu$) component of a gauge potential matrix.
 \begin{equation}
\Gamma^{\mu}_{\alpha \nu} = (\mathcal{A}_\alpha)^\mu_{~\nu}
\label{foura1}
\end{equation}
All quantities are functions of the d-dimensional coordinate $ x^\mu$. Next consider
a new coordinate system $\bar{x}^\mu (x)$. The conventional formula relating the
connection in the new coordinate system,
$\bar{\Gamma}^{\mu}_{\alpha \nu}$, to that in the original coordinates is of course
familiar. \cite{cosmograv} For our purposes, observe that the transformation formula
can be presented with the definition (4.A.1) as 
\begin{equation}
(\bar{\mathcal{A}}_\alpha ({\bar{x}}))^\mu_{\ \nu} = \left\{ (\mathcal{U}^{-1})^\mu
{_\rho} (\mathcal{A}_\beta (x))^\rho_{\ \sigma} (\mathcal{U})^\sigma_{\ \nu} + 
(\mathcal{U}^{-1})^\mu_{\ \sigma} \,
\frac{\partial}{\partial x^\beta} \, (\mathcal{U})^\sigma_{\ \nu} \right\} \,
\frac{\partial x^\beta}{\partial \bar{x}^\alpha}
\label{foura2}
\end{equation}
where
\begin{equation}
(\mathcal{U})^\mu_{\ \nu} = \frac{\partial x^\mu}{\partial \bar{x}^\nu} \quad, \quad
(\mathcal{U}^{-1})^\nu_{\ \mu} =  \frac{\partial \bar{x}^\nu}{\partial  x^\mu} 
\label{foura3}
\end{equation}
Thus we see that the matrix field $\mathcal{A}_\beta (x)$ transforms by $\frac{\partial x^\beta}{\partial
\bar{x}^\alpha}$ in its vector index, and also undergoes a gauge transformation by
the gauge function $\mathcal{U}$ in its matrix indices. In matrix notation
\begin{equation}
\mathcal{\bar{A}}_\alpha (\bar{x}) = \left\{\mathcal{U}^{-1} \mathcal{A}_\beta (x) \,
\mathcal{U} +
\mathcal{U}^{-1}
\frac{\partial}{\partial x^\beta}
\, \mathcal{U} \right\} \frac{\partial x^\beta}{\partial \bar{x}^\alpha}
\label{foura4}
\end{equation}
 If $\mathcal{U}$ were arbitrary, we would conclude from (\ref{foura4}) that $\mathcal{A}_\alpha$ is a
connection for the $G L(d, R)$ group. But the requirement that Christoffel connections be symmetric
in their lower indices (no torsion) is met only for $\mathcal{U}$ in the form (\ref{foura3}). (Here and
below gauge equivalents to geometrical objects are denoted by scripted letters. Coordinate
covariant derivatives acting on geometrical quantities are denoted by $D$; in
corresponding formulas where the derivative acts on gauge theory variables we
use as before $\mathcal{D}$.)

The Christoffel connection's notational analogy to a gauge potential
continues for covariant derivatives. A contravariant vector
behaves as a left-transforming group object.
\begin{subequations}
\begin{equation}
D_\alpha V^\mu = \partial_\alpha V^\mu + \Gamma^{\mu}_{\alpha \nu} V^\nu =
\partial_\alpha V^\mu + (\mathcal{A}_\alpha)^\mu_{\ \nu} V^\nu \sim
\mathcal{D}_\alpha
\mathcal{V}_L
\label{foura5a}
\end{equation}
while a covariant vector is right-transforming.
\begin{equation}
D_\alpha V_\mu = \partial_\alpha V_\mu - \Gamma^{\nu}_{\alpha \mu}
\, V_\nu = \partial_\alpha V_\mu - V_\nu (\mathcal{A}_\alpha)^\nu
{_\mu} \sim \mathcal{D}_\alpha \mathcal{V}_R 
\label{foura5b}
\end{equation}
Thus a mixed, second rank tensor is a matrix transforming in the
adjoint representation.
\begin{eqnarray}
D_\alpha V^\mu {_\nu} &=& \partial_\alpha V^\mu {_\nu} +
\Gamma^{\mu}_{\alpha \sigma} V^\sigma {_\nu} -
\Gamma^{\sigma}_{\alpha \nu} V^\nu {_\sigma} \nonumber\\
&=& \partial_\alpha V^\mu {_\nu} +
(\mathcal{A}_\alpha)^\mu_{\ \sigma} V^\sigma {_\nu} - V^\mu
{_\sigma} (\mathcal{A}_\alpha)^\sigma {_\nu} \nonumber\\
&\sim& \partial_\alpha \mathcal{V}_{\tiny \mbox{ADJ}} +
[\mathcal{A}_\alpha, \mathcal{V}_{\tiny \mbox{ADJ}}] \sim
\mathcal{D}_\alpha \mathcal{V}_{\tiny \mbox{ADJ}}
\label{foura5c}
\end{eqnarray}
\end{subequations}
The conventional formula in terms of Christoffel connections for
the Riemann curvature tensor translates to the gauge field
strength.
\begin{eqnarray}
R^\mu {_\nu} {_\alpha} {_\beta} \, (\Gamma) = (\mathcal{F}_{\alpha \beta})^\mu
{_\nu} = \partial_\alpha (\mathcal{A}_\beta)^\mu {_\nu} -
\partial_\beta (\mathcal{A}_\alpha)^\mu {_\nu} 
+[\mathcal{A}_\alpha, \, \mathcal{A}_\beta]^\mu {_\nu}
\label{foura6}
\end{eqnarray}
The Bianchi identity for the curvature and commutators of
covariant derivatives also translate freely.

Another useful formula relates the above to the {\it Vielbeine}
$e^a_\mu$ and spin connections $\omega^a_{\alpha b}$. Beginning
with
\begin{subequations}
\begin{equation}
D_\alpha \, e^a_{\mu} = \partial_\alpha e^a_{\mu} - \Gamma^{
\nu}_{\alpha \mu} \, e^a_\nu + \omega^a_{\alpha b}\ e^b_\mu =0,
\label{foura7a}
\end{equation}
we contract with the {\it Vielbein} inverse $e^\sigma_{a}$ to find
\begin{equation}
\Gamma^{\sigma}_{\alpha \mu} = e^\sigma_a\ \omega^a_{\alpha  b} e^b_{
\mu} + e^\sigma_{a} \partial_\alpha e^a_\mu.
\label{foura7b}
\end{equation}
\end{subequations}
Thus upon defining a matrix $\mathcal{E}$ with indices $(a, \mu)$
\begin{equation}
(\mathcal{E})^a_{\ \mu} = e^a_{\mu}, \ \ (\mathcal{E}^{-1})^\sigma \, _a = \,
e^\sigma_a,
\label{foura8}
\end{equation}
and viewing $\omega^a_{\alpha  b}$ as a matrix gauge
potential with indices $(a, b)$
\begin{equation}
\omega^{a}_{\alpha  b} = (\mathfrak{A}_\alpha)^a_{\ b},
\label{foura9}
\end{equation}
Eq. (4.A.7) reads, in matrix notation,
\begin{equation}
\mathcal{A}_\alpha = \mathcal{E}^{-1} \, \mathfrak{A}_\alpha\ \mathcal{E}
+
\mathcal{E}^{-1} \partial_\alpha \ \mathcal{E}.
\label{foura10}
\end{equation}
We see that the spin connection and the Christoffel connection are
gauge equivalent, with the {\it Vielbein} taking the role of the gauge
transformation function. This has the consequence that (gauge)
covariant quantities (like the Riemann tensor) can be constructed
either with Christoffel or spin connections, and the two expressions
are related by the covariant gauge transformation
$(\mathcal{E}^{-1}... \,\mathcal{E})$ built from the {\it Vielbein}. Thus for
example,
\begin{equation}
R^\mu_{\ \nu  \alpha \beta} (\Gamma) = e^\mu_a\, R^a_{\ b 
\alpha
\beta}\, (\omega) e^b_\nu
\label{foura11}
\end{equation}
where $R^a_{\ b 
\alpha
\beta}\, (\omega)$ is the Riemann curvature constructed in a familiar way
from the spin connection.

I have not been able to extend the gauge analogy any further.
Quantities arising in gravitational dynamics -- the Ricci
tensor and scalar -- involve contracting ``space-time" indices with
``gauge" indices. This requires using the metric tensor or the {\it Vielbeine}
(recall $g_{\alpha \beta} = e^a_\alpha \, e^b_\beta \, \eta_{a b}$). But
the former is not present in the gauge formalism, while the latter
plays the role of ``gauge" transformation functions, see
(\ref{foura8})-(\ref{foura11}). Nevertheless, even the limited analogy can be put
to good use.

 \setcounter{equation}{0}
 \def\theequation{4.B.\arabic{equation}}
\subsection*{B. Gravitational Chern-Simons term from gauge
theory Chern-Simons term}
We know how to construct a gauge theoretic Chern-Simons term in
three dimensions. Using the gravity-gauge theory dictionary,
specifically (\ref{one9}) and (\ref{foura1}), leads immediately to the formula
for the gravitational Chern-Simons term appropriate either to (2+1)-dimensional
space-time or to 3-dimensional space. \cite{3dmass}
\begin{equation}
W(\Gamma) = \frac{1}{4\pi^2} \int d^3 x \, \varepsilon^{\alpha \beta \gamma}
\bigg(\frac{1}{2}
\, \Gamma^\mu_{\alpha \nu} \partial_\beta \, \Gamma^\nu_{\gamma \mu} +
\frac{1}{3}
\Gamma^\mu_{\alpha 
\nu} \Gamma^\nu_{\beta  \omega} \Gamma^\omega_{\gamma 
\mu} \bigg)
\label{fourb1}
\end{equation}
Under coordinate transformations, $\Gamma (\sim \mathcal{A})$ transforms
according to (\ref{foura2})/(\ref{foura4}). Therefore from (\ref{one11}) and (\ref{one12}) it follows that
\begin{equation}
W(\bar{\Gamma}) = W(\Gamma) - \frac{1}{2 4\pi^2} \int d^3 x \, \varepsilon^{\alpha
\beta \gamma} <
\mathcal{U}^{-1} \partial_\alpha \, \mathcal{U} \,  \mathcal{U}^{-1} \partial_\beta \, \mathcal{U} \,
\mathcal{U}^{-1} \, \partial_\gamma \mathcal{U}>,
\label{fourb2}
\end{equation}
where $\mathcal{U}$ is the coordinate transformation matrix (\ref{foura3}) and the
last term in (\ref{fourb2}) is its winding number. We can restrict these transformations to
be sufficiently well-behaved so that there is no winding. Then $W(\Gamma)$ is a
coordinate invariant.

The Christoffel connection is constructed from the metric tensor $g_{\mu \nu}$. An
interesting geometrical quantity emerges when $W(\Gamma)$ is varied with respect
to $g_{\mu \nu}$. This is carried out in two steps: first vary (\ref{fourb1}) with respect to
$\Gamma$, and use (\ref{one10a}) (in tensor rotation) as well as the dictionary (\ref{foura1}) and
(\ref{foura6}). Then vary $\Gamma$ according to
\begin{equation}
\delta \Gamma^{\mu}_{\alpha  \nu} = \frac{g^{\mu \sigma}}{2} \bigg(D_\alpha
\delta g_{\sigma \nu} + D_\nu \delta g_{\sigma \alpha} - D_\sigma \delta g_{\alpha
\nu}\bigg).
\label{fourb3}
\end{equation}
Partial integration and the Bianchi identity leave
\begin{equation}
\delta W (\Gamma) \equiv -\frac{1}{4\pi^2}\int d^3 x \delta g_{\mu \nu}\, \sqrt{g}\,
C^{\mu
\nu},
\label{fourb4}
\end{equation}
where $C^{\mu \nu}$, called the Cotton tensor, reads
\begin{equation}
C^{\mu \nu} = \frac{1}{2\sqrt{g}}\ (\varepsilon^{\mu \alpha \beta} D_\alpha
R^\nu_\beta +
\varep D_\alpha R^\mu_\beta).
\label{fourb5}
\end{equation}
$C^{\mu \nu}$ is like a covariant curl of the Ricci tensor $R^\nu_{\beta} = R^{\mu
\nu}_{\ \ \mu \beta}$. [Instead of $R^\nu_{\beta}$ one can equivalently write in
(\ref{fourb5}) the Einstein tensor  $G^\nu_\beta \equiv R^\nu_\beta - \frac{1}{2}
\, \delta^\nu_\beta \, R^\mu_\mu$.]

$C^{\mu \nu}$ is obviously symmetric and traceless. Because it arises from the
variation of the invariant $W(\Gamma)$, $C^{\mu \nu}$ is also covariantly
conserved, as can be verified explicitly.

The role of $C^{\mu \nu}$ in 3-dimensional geometry is the following. In four or more
dimensions  there exists the Weyl tensor, which functions in two ways. It is the
non-Ricci part of the Riemann tensor; also it serves as a template for conformal
flatness: vanishing if and only if the space is conformally flat. In three dimensions
the Riemann tensor only has a Ricci part; the Weyl tensor is absent. [That is why
in 3-dimensional Einstein theory there is no curvature external to matter sources,
and  there are no propagating excitations.]

In the absence of the 3-dimensional Weyl tensor, $C^{\mu \nu}$ replaces it as the 
conformal template, vanishing if and only if space-time is conformally
flat. While these geometric properties of $C^{\mu \nu}$ are ancient knowledge, the
fact that it arises from varying the gravitational Chern-Simons term was a new
discovery, made possible by the gauge theory-gravity connection. \cite{3dmass}

The gauge theory Chern-Simons term can be added to a 3-dimensional Yang-Mills
theory, giving rise to massive but gauge invariant excitations. So also the
gravitational Chern-Simons term can supplement 3-dimensional Einstein theory,
where it has even more profound consequences. It converts a theory with no
propagating excitations into one with a massive propagating mode, all the time
maintaining diffeomorphism invariance! The equation of motion with an
energy-momentum tensor source reads
\begin{equation}
G^{\mu \nu} + \frac{1}{4 \pi^2 m}\ C^{\mu \nu} = -8\pi \,  G T^{\mu \nu}.
\end{equation}
Since $C^{\mu \nu}$ is of one higher derivative order than $G^{\mu \nu}$, the
strength parameter $m$ has dimension of mass, and $m$ is also the mass of the
propagating degree of freedom in the linearized theory. Thus in the Einstein limit,
cooresponding to $m \to \infty$, the massive excitation decouples leaving
non-propagating degrees of freedom. The quantum theory is well behaved, in spite of
the higher derivatives in $C^{\mu \nu}$. Unlike in the gauge theory, coupling
constant quantization is not needed, because the gravitational Chern-Simons term is
invariant against (well behaved) coordinate transformations.

The gravitational Chern-Simons term $W(\Gamma)$ may also be presented in terms
of the spin connection $W (\omega)$. From (\ref{one11}), (\ref{foura7b})/(\ref{foura10}) and
(\ref{fourb1}) follows
\begin{equation}
W(\Gamma) =W(\omega) \ - \frac{1}{2 4 \pi^2} \ \int d^3 x \varepsilon^{\alpha \beta
\gamma} <\mathcal{E}^{-1}
\,
\partial_\alpha
\, \mathcal{E}
\, \mathcal{E}^{-1}
\, \partial_\beta \, \mathcal{E} \, \mathcal{E}^{-1} \partial_\gamma \, \mathcal{E}>.
\label{fourb7}
\end{equation}
 Although the second term, being the winding number of $\mathcal{E}$, does not
contribute to the equations of motion, it should not be dropped; otherwise confusion
arises:
$W(\omega)$ in Minkowski space is a Chern-Simons term based on the local Lorentz
group SO(2.1), which does not support gauge transformations with non-zero
winding;
$W(\omega)$ is SO(2.1) invariant. However, it is usually belived that field theory may
be analytically continued to Euclidean space. In that case $W(\omega)$ is the
Chern-Simons term for SO(3) and this does possess non-trivial windings. So it is
difficult to understand how $W(\omega)$ could function in a quantum theory: with
Lorentzian signature its coefficent can be arbitrary, with Euclidean signature it must
be quantized! Which should it be?

Fortunately we need not answer this question, because $W(\omega)$ does not stand
alone: It is supplemented by the last term in (\ref{fourb7}) whose response to Euclidean
rotations must cancel any non-invariance of $W(\omega)$, because together they
equal
$W(\Gamma)$, which does not feel local rotations. 

Similarly to the work by Carroll {\it
et al.} on Chern-Simons extended electromagnetism, Pi and I added the gravitational
Chern-Simons term to 4-dimensional Einstein relativity. \cite{jackiwpi} Linearized
analysis indicates that, contrary to the electromagnetic case, wave propagation is not
affected. Another noteworthy feature is that vacuum space-times, which satisfy the
modified equations, necessarily possess vanishing gravitational Chern-Pontryagin
density, $\frac{1}{2}\, \varepsilon^{\alpha \beta \gamma \delta} R^\mu_{\ \nu \alpha
\beta} R^\nu_{\ \mu \gamma \delta} = 0$. Correspondingly, the Schwarzschild
space-time remains a solution, but Kerr space-time becomes modified.

  \setcounter{equation}{0}
 \def\theequation{4.C.\arabic{equation}}

\section*{C. Coordinate transformations in general relativity and gauge theory}
 \begin{description}
 \item{(i)}
 {\bf Response to changes in coordinates }
 \end{description}
In general, fields respond to an infinitesimal coordinate transformation, generated by the vector
$f$, through the Lie derivative $L_f$ with respect to $f$.
 \begin{eqnarray}
\delta_f x^\mu = -f^\mu (x) \label{fourc1} \\
\delta_f \, (\mbox{field}) = L_f  \, (\mbox{field})
\label{fourc2}
\end{eqnarray}
$L_f$ involves ordinary derivatives; for example for a covariant vector the action of
$L_f$ is
\begin{subequations}
\begin{equation}
\delta_f V_\alpha = L_f  \, V_\alpha = f^\mu \, \partial_\mu \, V_\alpha +
\partial_\alpha\,  f^\mu \, V_\mu
\label{fourc3a}
\end{equation}
while a contravariant vector responds by
\begin{equation}
\delta_f \, V^\alpha = L_f \, V^\alpha = f^\mu \, \partial_\mu \, V^\alpha - \partial_\mu \, f^\alpha \, V^\mu
\label{fourc3b}
\end{equation}
\end{subequations}
Moreover, when the space possesses a metric structure Lie derivatives of covariant objects (scalars, vectors, tensors etc.) remain covariant;
{\it i.e} replacing ordinary derivatives by coordinate covariant derivatives produces
no change in the formulas.

This is not true in gauge theories. For example the Lie derivative of the field strength
\begin{equation}
L_f F_{\alpha \beta} = f^\mu \partial_\mu\, F_{\alpha \beta} + \partial_{\alpha}\, 
f^\mu\, F_{\mu \beta} + \partial_\beta \, f^\mu \, \, F_{\alpha \mu}
\label{fourc4}
\end{equation}
is not gauge covariant because the derivative acting on $\, F_{\alpha \beta}$ is not
gauge covariant. Similarly, the Lie derivative of a vector potential, which follows
formulas (4.C.3) ($V \to A$) is not gauge covariant. Consequently coordinate
transformations in a gauge theory, implemented by Lie derivatives, loose gauge
covariance. 

We now ask: Is it possible to modify the implementation of coordinate
transformations in  gauge theories so that gauge covariance is preserved? The
positive answer that I  gave \cite{conformal} draws on notational analogies between
gauge and gravity fields. 

Let us record the gravity formulas. Under (4.C.1) and (4.C.2) the metric tensor transforms as
\begin{equation}
\delta_f \, g_{\mu \nu} = L_f \, g_{\mu \nu}\, = \, f^\alpha \, \partial_\alpha \, g_{\mu
\nu} + \partial_\mu\, f^\alpha\, g_{\alpha \nu} + \partial_\nu \, f^\alpha g_{\mu
\alpha} = D_\mu f_\nu + D_\nu f_\mu
\label{fourc5}
\end{equation}
The last equality follows from the previous when ordinary derivatives are replaced
by covariant derivatives, which also annihilate $g_{\mu \nu}$.\,With (\ref{fourc5}) we find
the response of the Christoffel connection from (\ref{fourb3}),
\begin{equation}
\delta_f\, \Gamma^\mu_{\alpha \nu} = f^\beta\, R^\mu_{\ \nu \beta \alpha} + D_\alpha (D_\nu\, f^\mu)
\label{fourc6}
\end{equation}
and furthermore
\begin{equation}
\delta_f R^\mu_{\ \nu \alpha \beta} = D_\alpha \delta\, \Gamma^\mu_{\beta \nu} -
D_\beta \delta\, \Gamma^\mu_{\alpha \nu} = L_f R^\mu_{\ \nu \alpha \beta}.
\end{equation}
\label{fourc7}
Both (\ref{fourc6}) and (\ref{fourc7}) exhibit a coordinate covariant response.

When we consider the gauge theoretic analog to (\ref{fourc6}), we expect to find that  the coordinate
transformation of the covariant vector potential, {\it i.e.}\,its Lie derivative, can be presented
analogously to (\ref{fourc6}) as the sum of two terms: a projection on the field strength and a total
derivative. This is indeed the case, as is readily established by adding and subtracting suitable terms in
(\ref{fourc3a})
$(V_\alpha \to A_\alpha)$.
\begin{eqnarray}
\delta_f \, A_\alpha  &=& f^\mu (\partial_\mu \, A_\alpha - \partial_\alpha \, A^\mu +
[A_\mu, A_\alpha]) \nonumber\\
                         &+&  f^\mu (\partial_\alpha \, A_\mu - [A_\mu, A_\alpha]) +
\partial_\alpha \, f^\mu \, A_\mu \nonumber\\
                         &=&  f^\mu \, F_{\mu \alpha} + \mathcal{D}_\alpha (f^\mu \, A_\mu) 
\label{fourc8}
\end{eqnarray}
It is the last term in (\ref{fourc8}) that spoils gauge covariance. We
recognize it as an infinitesimal gauge transformation with gauge function of
$f^\mu
\, A_\mu$. But in a gauge theory, gauge transformations can be performed at will.
We use this freedom to redefine the response of gauge variables by supplementing
the Lie derivatives with the gauge transformation that removes the last term in
(\ref{fourc8}). Thus the modified, but gauge equivalent response reads
\begin{equation}
\bar{\delta}_f \, A_\alpha = f^\beta F_{\beta \alpha}
\label{fourc9}
\end{equation}
and (\ref{fourc9}) has the consequence that
\begin{eqnarray}
\bar{\delta}_f  F_{\alpha \beta} &=& \mathcal{D}_\alpha \, \bar{\delta}_f \, A_\beta - 
\mathcal{D}_\beta\, \bar{\delta}_f A_\alpha \nonumber\\ &=& f^\mu\,
\mathcal{D}_\mu \, F_{\alpha \beta} + \partial_\alpha \, f^\mu \, F_{\mu \beta} +
\partial_\beta f^\mu F_{\alpha \mu} 
\label{fourc10}
\end{eqnarray}
This is gauge covariant and differs from the usual formula (\ref{fourc4}) by a gauge transformation of
$F_{\alpha \beta} $ generated by $f^\mu A_\mu$. We may view (\ref{fourc10}) as defining a
gauge covariant Lie derivative.

We thus achieve the goal of describing coordinate transformation of gauge fields in a
gauge covariant manner, with a gauge covariant Lie derivative. But there is a price
to pay: The gauge covariant Lie derivatives follow a closure rule that differs from
conventional Lie derivatives. The commutator of two gauge covariant Lie
derivatives, with respect to two vectors, $f$ and $g$, closes on the gauge covariant
Lie derivative with respect to the Lie bracket of $f$ and $g$ plus a gauge
transformation generated by
$f^{\alpha} \, g^\beta F_{\alpha \beta}$. 
\begin{description}
\item{(ii)} \vspace{-16pt}
{\bf Invariant fields and constants of motion}
\end{description}
To determine whether a  generic field is invariant against a coordinate
transformation generated by $f$, we check whether its Lie derivative annihilates
$\phi$.
\begin{equation}
L_f \phi = 0 \ \Rightarrow (\mbox{invariant field} \, \phi)
\label{fourc11}
\end{equation}
For the metric tensor this is the statement of the Killing equation, see (\ref{fourc5}).
\begin{equation}
\delta_f g_{\mu \nu} = L_f \, g_{\mu \nu} = D_\mu \, f_\nu + D_\nu \, f_\mu =0
\label{fourc12}
\end{equation}
From (\ref{fourc6}) and (4.C.7) immediately follow well known conditions on Killing vectors.
\cite{inchoate}
\begin{eqnarray}
f^\beta \, R^\mu_{\ \nu \beta \alpha} = -D_\alpha (D_\nu f^\mu) \label{fourc13} \\
L_f \, R^\mu_{\ \nu \alpha \beta} = 0
\label{fourc14}
\end{eqnarray}

But in a gauge theory a  condition weaker than (\ref{fourc11}) is appropriate: A coordinate
invariant configuration in  a gauge theory need not be annihilated  by the Lie
derivative, rather a gauge transformation may survive. In other words, a
gauge field configuration should still be considered as invariant, if any non invariance
can be compensated by a gauge transformation. Applying this condition to the gauge
covariant transformation law (\ref{fourc9}), there emerges a gauge covariant criterion for
an invariant gauge field configurations.
\begin{equation}
f^\beta F_{\beta \alpha} = \mathcal{D}_\alpha \Phi_f \Rightarrow \, \mbox{invariant
gauge field} \, F_{\alpha \beta}
\label{fourc15}
\end{equation}
Here $\Phi_f$ is an unspecified quantity, linear in $f$. This is the gauge theoretic
analog to (\ref{fourc13}), except that in the gravity formula the quantity corresponding
to $\Phi_f$ is specified explicitly as $-D_\nu \, f^\mu$.

Manton and I showed that $\Phi_f$ generates the gauge transformation needed to
compensate any coordinate asymmetry in an invariant gauge field
configuration.\,Therefore $\Phi_f$ also contributes to the conserved constant of
motion, which characterizes motion in the presence of such an invariant gauge
field. \cite{nmanton}

This is the origin of the celebrated addition to the angular momentum in the field of a Dirac magnetic monopole: 
The conserved angular momentum comprises the kinematical angular momentum
supplemented by the radial unit vector, which is also present in (\ref{fourc15}), when
$F_{\beta \alpha}$ is the magnetic monopole field and $f$ generates spatial rotations
around the axis ${\bf a}: f^\mu = (0, {\bf r} \times {\bf a})$, then
$\Phi_f = \hat{\bf r} \cdot \bf a$ Similarly, the `tHooft-Polyakov magnetic monopole is
spherically symmetric up to an isospin gauge transformation. The angular
momentum constant of motion therefore acquires an isospin component
$\bf T$ in addition to the kinematical angular momentum $\bf L$.
\begin{equation}
{\bf J} = {\bf L} + {\bf T}
\label{fourc16}
\end{equation}
This is the origin of the previously mentioned conversion of isospin to spin.

\raggedright


\begin{thebibliography}{99}
\bibitem{cnyang}
C.-N. Yang and R.L. Mills, ``Conservation of Isotopic Spin and Isotopic Gauge
Invariance" Phys. Rev. {\bf 96}, 191 (1954).

\bibitem{thooft}
G. `tHooft, `` The Renomalization Procedure for Yang-Mills Fields", Ph.D Thesis, Utrecht
University (1972);
G. `tHooft and M. Veltman, ``Regularization and Renormalization of Gauge Fields" Nucl
Phys. {\bf B44}, 189 (1972).

\bibitem{jsbell}
J.S. Bell and R. Jackiw, ``A PCAC Puzzle: $\pi^0 \to 2 \gamma$ in the $\sigma$-model"
Nuovo Cim. {\bf A60}, 47 (1969).

\bibitem{fumiy}
H. Fukuda and Y Miyamoto, ``On the $\gamma$-Decay of  Neutral Meson" Prog.
Theoret. Phys. {\bf 4}, 347 (1949);
J. Steinberger, ``On the Use of Subtraction Fields and the Lifetimes of some Types of
Meson Decay" Phys. Rev. {\bf 76}, 1180 (1949);
J. Schwinger, ``Gauge Invariance and Vacuum Polarization" Phys. Rev. {\bf 82}, 664
(1951);
S. Adler,  ``Axial Vector Vertex in Spinor Electrodynamics" Phys. Rev. {\bf 177},
2426 (1969).

\bibitem{dsuther}
D. Sutherland, ``Current Algebra and some Nonstrong Mesonic Decays" Nucl. Phys.
{\bf B2}, 433 (1967);
M. Veltman, ``Theoretical Aspects of High Energy Neutrino Interactions" Proc. Roy.
Soc. London {\bf A 301}, 107 (1967).

\bibitem{dgjack}
D. Gross and R. Jackiw, ``Effect of Anomalies on Quasirenormalizable Theories" Phys.
Rev. D {\bf 6}, 477 (1972).

\bibitem{cbouch}
C. Bouchiat, J. Iliopoulos and P. Meyer, ``An Anomaly  free Version of Weinberg's
Model" Phys. Lett. {\bf B38}, 519 (1972).

\bibitem{abelvin}
A. Belavin, A. Polyakov, A. Schwartz and Y. Tyupkin, ``Pseudoparticle Solutions of the
Yang-Mills Equations" Phys. Lett. {\bf B59}, 85 (1975).

\bibitem{gthooft}
G. `tHooft, ``Symmetry Breaking Through Bell-Jackiw Anomalies" Phys. Rev. Lett. {\bf
37}, 8 (1976).

\bibitem{crebbi}
R. Jackiw and C. Rebbi, ``Vacuum Periodicity in a Yang-Mills Quantum Theory" Phys.
Rev. Lett. {\bf 37}, 172 (1976).

\bibitem{crdgross}
C. Callan, R. Dashen and D. Gross, ``The Structure of the Gauge Theory Vacuum" Phys.
Lett. {\bf B63}, 334 (1976).

\bibitem{rjebbi}
R. Jackiw and C. Rebbi, ``Conformal Properties of a Yang-Mills Pseudoparticle" Phys.
Rev D {\bf 14}, 517 (1976).

\bibitem{rjebbi2}
R. Jackiw and C. Rebbi, ``Degrees of Freedom in Pseudoparticle Systems" Phys. Lett.
{\bf 67}, 189 (1977).

\bibitem{aschwartz}
A. Schwartz, ``On Regular Solutions of Euclidean Yang-Mills Equations" Phys. Lett. {\bf
67B}, 172 (1977);
M. Atiyah, N.Hitchin and I. Singer, ``Deformations of Instantons" Proc. Nat. Acad. Sci.
{\bf 74}, 2662 (1977).

\bibitem{cnohl}
R. Jackiw, C. Nohl and C. Rebbi, ``Conformal Properties of Pseudoparticle
Configurations" Phys. Rev. D {\bf 15}, 1642 (1977).

\bibitem{aityah}
M. Aityah, N. Hitchen, V. Drinfeld and Y. Manin, ``Construction of Instantons" Phys.
Lett. {\bf A65}, 185 (1978).

\bibitem{spinor}
R. Jackiw and C. Rebbi, ``Spinor Analysis of Yang-Mills Theory" Phys. Rev. D {\bf 16},
1052 (1977).

\bibitem{magmono}
G. `tHooft, ``Magnetic Monopoles in Unified Gauge Theories" Nucl. Phys. {\bf 79}, 276
(1974);
A. Polyakov, ``Particle Spectrum in the Quantum Field Theory" Zh. Eksp. Teor. Fiz. Pis'
ma Red. {\bf 20}, 430 (1974) [English translation: JETP Lett. {\bf 20}, 430 (1974)].

\bibitem{isospin}
R. Jackiw and C. Rebbi, ``Spin from Isospin in a Gauge Theory" Phys. Rev. Lett. {\bf
36}, 1116 (1976).

\bibitem{hasenfratz}
P. Hasenfratz and G. `tHooft, ``A Fermion-Boson Puzzle in a Gauge Theory" Phys. Rev.
Lett. {\bf 36}, 1119 (1976).

\bibitem{soliton21}
R. Jackiw and C. Rebbi, ``Solitons with Fermion Number $\frac{1}{2}$" Phys. Rev. D {\bf
13}, 3398 (1976);
R. Jackiw and J. R. Schrieffer, ``Solitons with Fermion Number $\frac{1}{2}$ in
Condensed Matter and Relativistic Field Theories" Nucl. Phys. {\bf B190}, 253 (1981).

\bibitem{ccallias}
C. Callias, ``Index Theorems on Open Spaces" Commun. Math. Phys. {\bf 62}, 213
(1978);\\ R. Bott and R. Seeley, ``Some Remarks on the Paper of Callias" Commun.
Math. Phys. {\bf 62}, 245 (1978).

\bibitem{rjackiw}
R. Jackiw, ``The Yang-Mills Vacuum as a Bloch Wave" APS Spring Meeting,
Washington D.C. (1977); reprinted in R. Jackiw, {\it Diverse Topics  in Theoretical and
Mathematical Physics} (World Scientific, Singapore, 1995).

\bibitem{3dmass}
S. Deser, R. Jackiw and S. Templeton, `` Three Dimensional Massive Gauge Theories"
Phys. Rev. Lett. {\bf 48}, 975 (1982); ``Topologically  Massive Gauge Theories" Ann.
Phys. {\bf 140}, 372 (1982),  (E) {\bf 185}, 406 (1988).

\bibitem{selfdual}
R. Jackiw and E Weinberg, ``Self-Dual Chern-Simons Vortices" Phys. Rev. Lett. {\bf
64}, 2234 (1990).

\bibitem{multivor}
J. Hong, Y. Kim and P.-Y. Pac, ``Multivortex Solitons of the  Abelian
Chern-Simons-Higgs Theory" Phys. Rev. Lett. {\bf 64} 2330 (1990).

\bibitem{pisoliton}
R. Jackiw and S.-Y. Pi, ``Soliton Solutions to the Gauged Nonlinear Schr\"{o}dinger
Equation on the Plane" Phys. Rev. Lett. {\bf 64}, 2969 (1990).

\bibitem{reviews}
For reviews see R. Jackiw and S.-Y. Pi, ``Self-Dual Chern-Simons Solitons" Prog. Theor.
Phys. (Kyoto) Suppl. {\bf 107}, 1 (1992); G. Dunne ``Self-Dual Chern-Simons
Theories" Lecture Notes in Physics {\bf m36} (Springer, Berlin, 1995).

\bibitem{nonvariance}
N. Redlich, ``Gauge Noninvariance and Parity Violation of Three-Dimensional
Fermions" Phys. Rev. Lett. {\bf 52}, 18 (1984).

\bibitem{holstein}
B. Holstein, ``Anomalies for Pedestrians" Am. J. Phys. {\bf 61}, 142 (1993).

\bibitem{dilatation}
S. Coleman and R. Jackiw, ``Why Dilatation Generators do not Generate Dilatations?"
Ann. Phys. {\bf 67}, 552 (1971).

\bibitem{parityvio}
S. Carroll, G. Field and R. Jackiw, ``Limits on a Lorentz and Parity Violating
Modification of Electrodynamics" Phys. Rev. D {\bf 41}, 1231 (1990);
S. Carroll and G. Field, ``Is there Evidence for Cosmic Anisotropy in the Polarization of
Distant Galaxies?" Phys. Rev. Lett. {\bf 79}, 2394 (1997).

\bibitem{gravtheor}
W. Bardeen and B. Zumino, ``Consistent and Covariant Anomalies in Gauge and
Gravitational Theories" Nucl. Phys. {\bf B244}, 421 (1984).

\bibitem{cosmograv}
Our definitions and conventions for geometrical entities follow S. Weinberg,
{\it Gravitation and Cosmology} (Wiley, New York NY, 1972), except that our
Riemann tensor is the negative of his, as is our Lorentzian metric tensor.

\bibitem{jackiwpi}
R. Jackiw and S.-Y. Pi, ``Chern-Simons Modification of Gravity"
Phys. Rev. D {\bf 68}, 104012 (2003).

\bibitem{conformal}
R. Jackiw, ``Gauge Covariant Conformal Transformations" Phys. Rev. Lett. {\bf 41},
1635 (1978); ``Invariance, Symmetry and Periodicity in Gauge Theories" Acta Phys.
Austr. Suppl. XXII, 383 (1980), reprinted in R. Jackiw, {\it Diverse Topics in
Theoretical and Mathematical Physics} (World Scientific, Singapore 1995).

\bibitem{inchoate}
Formula (4.C.14) is equivalent to the inchoate (13.1.12)  in
Weinberg \cite{cosmograv}.

\bibitem{nmanton}
R. Jackiw and N. Manton, ``Symmetries and Conservation Laws in Gauge Theories"
Ann. Phys. {\bf 127}, 257 (1980).
\end{thebibliography}
\end{document}